\documentclass[]{IAC_style}
\usepackage{multirow}
\usepackage{float}
\begin{document}

\IACpaperyear{2025} 
\IACpapernumber{76th International Astronautical Congress 2025 Paper-IAC-25,A6,11,10,x97872} 
\IAClocation{Sydney, Australia} 
\IACdate{28-3 October 2025} 

\title{Micro-Doppler signatures and object characterisation of space debris with radio telescopes}

\IACauthor{Dr. Guifre Molera Calves$^{a,}$}{2}{1}
\IACauthor{Dr. Shinji Horiuchi}{3}{0}
\IACauthor{Dr. Edwin Peters}{4}{0}
\IACauthor{Dr. Ed Kruzins}{4}{0}
\IACauthor{Dr. Nick Stacy}{5}{0}

\IACauthoraffiliation{University of Tasmania, Australia \normalfont{E-mail:~\authormail{guifre.moleracalves@utas.edu.au}}}
\IACauthoraffiliation{HENSOLDT Australia, Australia}
\IACauthoraffiliation{Commonwealth Scientific and Industrial Research Organisation (CSIRO), Australia}
\IACauthoraffiliation{University of New South Wales, Australia}
\IACauthoraffiliation{Blue Moon Lab, Australia}

\abstract{
This study presents a novel multi-static radar technique for space debris characterisation using micro-Doppler signatures, developed within the Southern Hemisphere Asteroid Radar Programme (SHARP). The method employs C-band continuous waveforms transmitted from NASA's Deep Space Network (DSN) Canberra station, with reflected signals captured by distributed ground-based telescopes converted from astronomical to radar operations. From March 2021 to September 2025, we conducted systematic observations of 20 distinct rocket bodies at various orbital altitudes and object sizes. Micro-Doppler analysis of axial rotation signatures enabled extraction of rotation periods, object dimensions, surface characteristics, and mass distribution parameters with enhanced accuracy in orbital refinement. We implemented advanced imaging reconstruction techniques to generate morphology data of the targets. Results demonstrate successful determination of debris rotation periods with second-level precision, dimensional estimates within 10\% accuracy compared to known specifications, and improved orbital parameter determination reducing position uncertainties by up to 30\%. The technique provides a cost-effective approach for space situational awareness, leveraging existing astronomical infrastructure to enhance Australia's sovereign capabilities in tracking both near-Earth objects and space debris. This multi-static configuration offers significant advantages over traditional monostatic radar systems for geostationary and highly elliptical orbit debris monitoring.
}

\maketitle
\thispagestyle{fancy}

\section*{Nomenclature}
\noindent SHARC = Southern Hemisphere Asteroid Research Consortium\\
SHARP = Southern Hemisphere Asteroid Radar Program\\
SG = Southern Guardian sensors \\
VLBI = Very Long Baseline Interfero-metry\\

\section{Introduction}

The proliferation of space debris in Earth's orbit has become one of the most pressing challenges facing modern space operations. The number of satellites launched in 2024 alone surpassed 2,800 objects, whilst no established mechanism exists to practically remove debris en masse~\cite{mark2019review}. This rapidly growing population of space debris poses significant collision risks to critical space-based infrastructure, making space domain awareness (SDA) increasingly critical for understanding precise orbits, spin state evolution, debris characterisation, and object recognition across all orbital regimes~\cite{holzinger2018challenges, murphy2018optimal}.

Traditional ground-based radar systems for space surveillance face significant limitations, particularly for objects in geosynchronous Earth orbit (GEO) and other high-altitude regimes. The extreme tracking distances of up to 36,000 km result in substantial sensitivity losses, requiring either prohibitively large apertures or impractical power levels for conventional monostatic radar configurations. Additionally, knowing object positions at detection time is insufficient, as spin states of debris objects can evolve significantly over extended periods~\cite{Benson_2023}. Multiple observations can monitor these changes and improve models for attitude-dependent debris perturbations.

The concept of using large parabolic radio telescopes for bistatic radar experiments has proven successful in planetary radar applications since the first asteroid detection with the Goldstone radio telescope in 1968~\cite{ostro1983planetary}. Over 1,000 near-Earth asteroids have since been observed at microwave frequencies, primarily using Deep Space Network (DSN) stations and large radio telescopes. This success has extended to the Southern Hemisphere, with experiments utilising DSN Tidbinbilla, the Murriyang antenna at Parkes Observatory, and the Australian Telescope Compact Array~\cite{benson2018, Horiuchi_2021, kruzins2023}. The University of Tasmania joined this domain with their network of large (30-m, 26-m) and small (12-m) radio telescopes~\cite{lovell2013,white2025development}, initially developed for very long baseline interferometry (VLBI) applications in geodesy and planetary science.

The adaptation of radio telescopes for space debris tracking offers several critical advantages. Radio telescopes typically possess large apertures with low system temperatures, often an order of magnitude lower than conventional radars. Their pointing and tracking capabilities are well-matched to GEO targets, whilst wide-band receivers provide flexibility to pair with multiple transmitting radars. Most importantly, their excellent frequency standards enable coherent bistatic observations, essential for characterisation through micro-Doppler analysis. The functionality required for radio telescopes to contribute to space domain awareness is remarkably similar to that required for planetary radar or spacecraft tracking, making this adaptation both practical and cost-effective~\cite{agaba2017system}.

Micro-Doppler signatures resulting from object rotation provide unique insights into debris characteristics that cannot be obtained through positional tracking alone. These signatures enable extraction of rotation periods, dimensional estimates, surface properties, and mass distribution parameters. For tumbling satellites and rocket bodies, Doppler-time intensity analysis and tomographic imaging techniques can generate resolved images even when targets are unresolved in range, particularly valuable for objects at high altitudes where bandwidth limitations prevent traditional high-resolution radar imaging~\cite{serrano2024}. The bistatic configuration offers additional advantages over monostatic systems, including improved geometric diversity and enhanced sensitivity through power-aperture product optimisation.

The Southern Hemisphere Asteroid Radar Programme (SHARP) represents a unique consortium that leverages Australia's strategic geographical position and advanced radio telescope infrastructure to address the critical gap in Southern Hemisphere space surveillance capabilities. SHARP provides unprecedented opportunities to characterise space debris across multiple orbital regimes by using distributed ground-based telescopes in conjunction with the DSN station. This approach not only enhances Australia's sovereign space situational awareness capabilities, but also contributes to global space safety through improved understanding of debris population characteristics and orbital evolution.

\begin{figure}[H]
\centering
\includegraphics[width=\columnwidth]{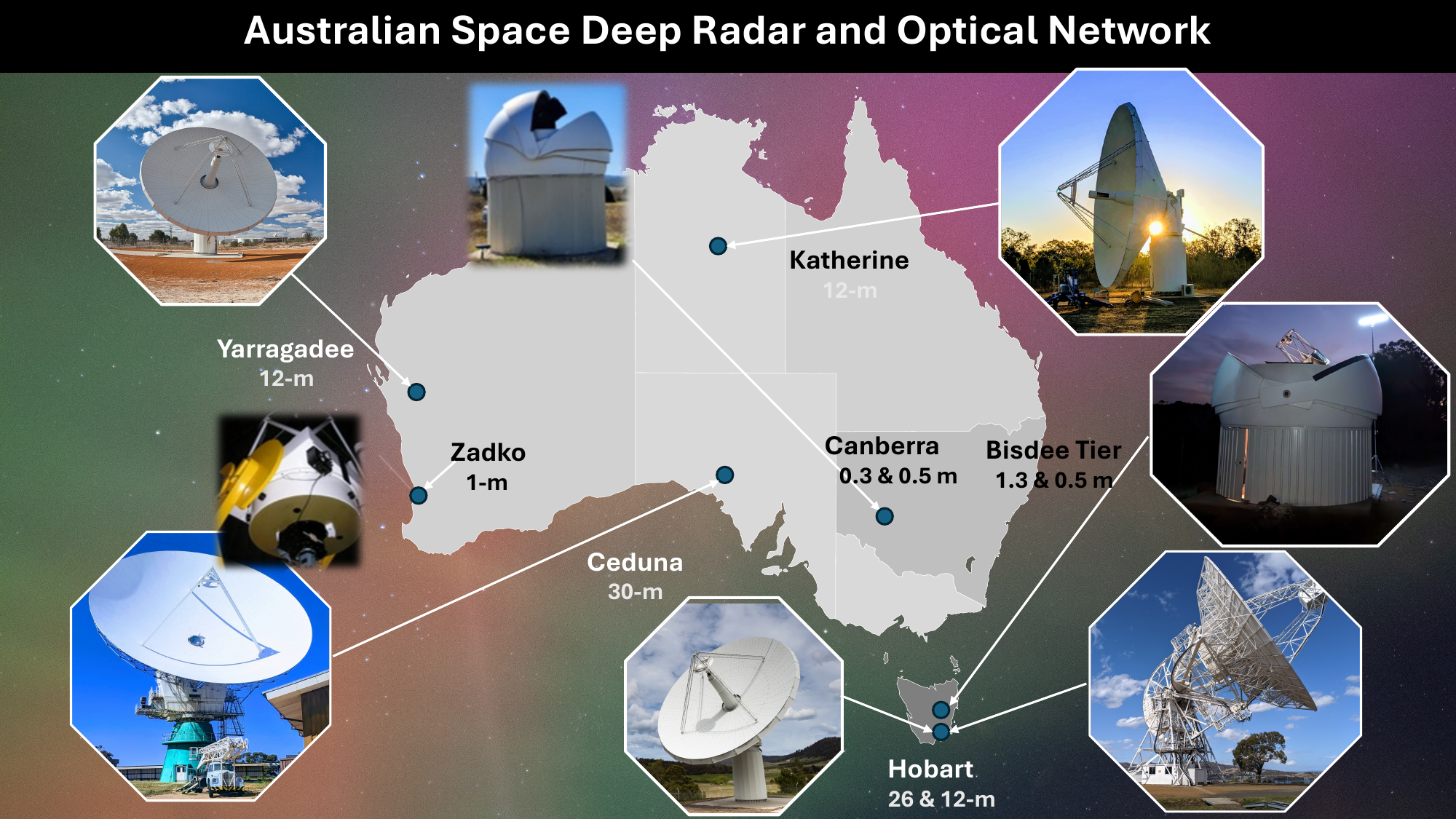}
\label{fig:sharp_network}
\caption{The SHARP network configuration showing the distributed antenna system used for bi-static radar observations of space debris.}
\end{figure}

This paper presents the results of systematic multi-static radar observations conducted from March 2021 to 2025, targeting space debris objects at various orbital altitudes ranging from low Earth orbit to geosynchronous regimes. We demonstrate the characterisation of 20 distinct rocket bodies and defunct satellites through micro-Doppler signature analysis, enabling the extraction of rotation periods, dimensional parameters, and surface characteristics with enhanced accuracy. The methodology employs C-band continuous waveforms transmitted from the DSN Canberra station, with reflected signals captured by a distributed network of radio telescopes that have been adapted from their traditional astronomical purposes to specialised radar operations.

The paper is organised as follows: Section~2 describes the experimental methodology and data processing techniques; Section~3 discusses the observations conducted and methodology, Section~4 presents the characterisation results for objects across different orbital altitudes; and Section~5 provides conclusions and recommendations for future work.

\section{Material and methods}

\subsection{Experimental Configuration}

The bi-static radar observations were conducted using NASA's Deep Space Network (DSN) Deep Space Stations (DSS) in Canberra (DSS-34, DSS-35 or DSS-36, 34-m) as the transmitting facility, operating at C-band frequencies (7.15 GHz) with continuous waveforms. The DSN antenna was configured to transmit signals at different polarisations—linear horizontal (LH), linear vertical (LV), and circular (RHC/LHC)—to optimise signal reflection characteristics for various target orientations and surface properties.

The receiving network comprised multiple distributed ground-based radio telescopes that were converted from astronomical to radar operations for these observations. The primary receiving stations included: the CSIRO Australian Compact Telescope Array (ATCA),  the University of Tasmania's Hobart 12-m antenna (Hb), Katherine 12-m antenna (Ke), Ceduna 30-m antenna (Cd), Yarragadee 12-m antenna (Yg), and Hobart 26-m antenna (Ho). This multi-static configuration provided enhanced geometric diversity and improved signal-to-noise ratios through power-aperture product optimisation.

\subsection{Radio Telescope Array Configuration}

The University of Tasmania operates a distributed array of radio telescopes strategically positioned across Australia to provide comprehensive coverage for space debris observations. This network, also known as Southern Guardian (SG), leverages existing astronomical infrastructure, adapting these facilities for bi-static radar operations through specialised instrumentation and signal processing capabilities. The array configuration enables simultaneous multi-station observations, providing enhanced geometric diversity and improved orbital determination accuracy through differential range and Doppler measurements.

\begin{figure}[H]
\centering
\includegraphics[width=\columnwidth]{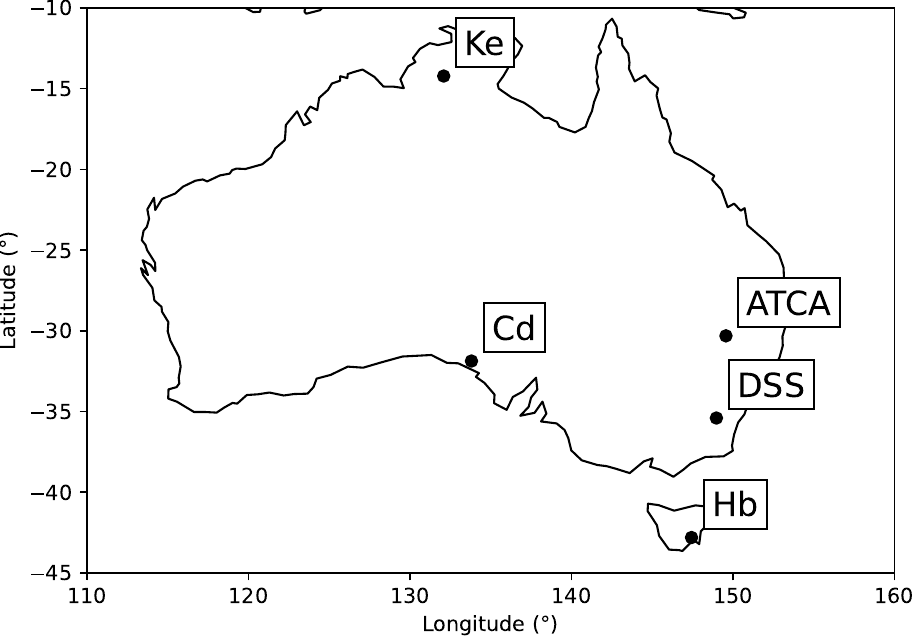}
\label{fig:utas_array}
\caption{Geographic distribution of the University of Tasmania radio telescope array used for space debris characterisation.}
\end{figure}

The UTAS network comprises strategically distributed radio telescopes across the Australian continent, including the Hobart, Katherine, Yarragadee, and Ceduna antennas. This geographical distribution enables diverse viewing geometries and enhanced orbital determination capabilities. The observations are complemented by NASA's Deep Space Network stations at Tidbinbilla near Canberra. Table~\ref{tab:antenna_specs} provides the complete specifications and precise coordinates for all participating stations in the SHARP network.

\begin{table}[H]
\centering
\caption{Radio telescope specifications and coordinates for SHARP network stations.}
\label{tab:antenna_specs}
\footnotesize
\begin{tabular}{lllll}
\hline
\textbf{Station} & \textbf{D} & \textbf{Lat.} & \textbf{Long.} & \textbf{Elev.} \\
\hline
\multicolumn{5}{l}{\textit{UTAS network}} \\
Hobart (Hb) & 26 & -42.80°S & 147.43°E & 41 \\
Katherine (Ke) & 12 & -14.46°S & 132.26°E & 145 \\
Yarragadee (Yg) & 12 & -29.04°S & 115.34°E & 244 \\
Ceduna (Cd) & 30 & -31.86°S & 133.81°E & 160 \\
\hline
\multicolumn{5}{l}{\textit{NASA DSN}} \\
DSS-34 & 34 & -35.40°S & 148.98°E & 684 \\
DSS-35 & 34 & -35.40°S & 148.98°E & 691 \\
DSS-36 & 34 & -35.40°S & 148.98°E & 686 \\
\hline
\multicolumn{5}{l}{\footnotesize Elevation and Diameter (D) are in meters}
\end{tabular}
\end{table}

\subsection{Signal Processing and Doppler Compensation}

Three distinct Doppler compensation strategies were systematically tested to optimise target detection and characterisation:

\begin{enumerate}
\item \textbf{No Doppler Compensation}: Raw received signals were processed without frequency corrections, allowing direct observation of natural Doppler shifts caused by target motion and rotation.

\item \textbf{Pre-Doppler Compensation for Stations}: Frequency corrections were applied to compensate for relative motion between transmitting and receiving stations, isolating target-induced Doppler signatures.

\item \textbf{Geocentric Doppler Compensation}: The pre-Doppler frequency correction was applied using the centre of the Earth as reference, the Doppler difference is only between the station surface and geocentre.
\end{enumerate}

\subsection{Observational Campaign}

From March 2021 to September 2025, a systematic observational campaign was conducted comprising of 14 dedicated observation sessions (designated sda001 through sda014). Each session targeted multiple objects simultaneously, with observation durations ranging from 2 to 8 hours depending on target visibility windows and orbital geometries. The list of sessions conducted since 2021 until 2025 are summarised in Table~\ref{tab:sessions_summary}. Some sessions including sda003, sda006, sda007 and sda008 were not successful due to technical or weather issues.

\begin{table*}[ht]
\centering
\caption{Summary of observation sessions conducted during the SHARP SDA campaign.}
\label{tab:sessions_summary}
\footnotesize
\begin{tabular}{llll}
\hline
\textbf{Session} & \textbf{Date} & \textbf{Antennas} & \textbf{Target Observed} \\
\hline
sda001 & 21 Mar 2021 & Cd, DSN, Hb, Ke & Atlas V R/B (40947) \\
\hline
\multirow{4}{*}{sda002} & \multirow{4}{*}{03 Jun 2021} & \multirow{4}{*}{DSN, Hb, Ke} & Atlas V R/B (40947) \\
& & & SL-8 R/B (16292) \\
& & & LCS-1 (1361) \\
& & & GOES-9 (23581) \\
\hline
\multirow{2}{*}{sda004} & \multirow{2}{*}{18 Jul 2021} & \multirow{2}{*}{Cd, DSN, Hb, Ho} & Moon \\
& & & AUSSAT-1 (15993) \\
\hline
\multirow{3}{*}{sda005} & \multirow{3}{*}{25 Aug 2021} & \multirow{3}{*}{DSN, Hb} & BREEZE-M R/B (39128) \\
& & & SL-12 R/B (24919) \\
& & & Ariane DEB (23845) \\
\hline
\multirow{7}{*}{sda009} & \multirow{7}{*}{18 May 2022} & \multirow{7}{*}{DSN, Hb, Ke} & SL-12 R/B (24919) \\
& & & CZ-3A R/B (41435) \\
& & & BREEZE-M R/B (39128) \\
& & & SL-12 R/B (37140) \\
& & & BLOCK DM-SL R/B (37817) \\
& & & SL-8 R/B (8897) \\
& & & PAGEOS DEB (2511) \\
\hline
\multirow{4}{*}{sda010} & \multirow{4}{*}{25 Aug 2022} & \multirow{4}{*}{DSN, Hb, Ke} & AUSSAT-1 (15994) \\
& & & GOES-12 (26871) \\
& & & LCS-1 (1361) \\
& & & Atlas V R/B (40947) \\
\hline
\multirow{2}{*}{sda011} & \multirow{2}{*}{26 Aug 2022} & \multirow{2}{*}{DSN, Hb, Ke} & Atlas V R/B (40947) \\
& & & CALSPHERE-1 (900) \\
\hline
\multirow{2}{*}{sda012} & \multirow{2}{*}{06 Dec 2023} & \multirow{2}{*}{Hb} & HALCA (24720) \\
& & & ARIANE DEB (37817) \\
\hline
\multirow{7}{*}{sda013} & \multirow{7}{*}{22 May 2024} & \multirow{7}{*}{Yg, DSN, Hb, Ke} & GOES-9 (23581) \\
& & & HGS-1 (25126) \\
& & & Atlas-5 R/B (45466) \\
& & & LCS-1 (1361) \\
& & & H-2A R/B (42918) \\
& & & Navstar-18 (20452) \\
& & & Moon \\
\hline
sda014 & 13 Jun 2024 & Cd, DSN, Hb, Ke & Moon \\
\hline
\end{tabular}
\end{table*}

A total of 20 distinct space debris objects were successfully characterised during the campaign, spanning a range of orbital altitudes from low Earth orbit (LEO) to geosynchronous Earth orbit (GEO). The primary targets included various rocket bodies, defunct satellites, and calibration spheres, each selected for their relevance to space debris research and their suitability for micro-Doppler analysis. Table~\ref{tab:objects_summary} summarises the key characteristics of these objects, including their NORAD IDs, types, altitudes, number of observation sessions, and overall status based on the data collected.

\begin{table*}[ht]
\centering
\caption{Summary of primary space debris objects characterised during the observational campaign.}
\label{tab:objects_summary}
\footnotesize
\begin{tabular}{llllll}
\hline
\textbf{Object} & \textbf{Object ID} & \textbf{NORAD} & \textbf{Type} & \textbf{Alt. (km)} & \textbf{Sessions} \\
\hline
Atlas V R/B & 2015-056B & 40947 & Rocket Body & 35,500 & 4 \\
Atlas V R/B & 2020-022C & 45466 & Rocket Body & 35,500 & 1 \\
AUSSAT-1 & 1985-076B & 15993 & Comms Sat. & 36,000 & 2 \\
BLOCK DM-SLR/B & 2011-051B & 37817 & Rocket Body & 35,500 & 1 \\
BREEZE-M R/B & 2013-014B & 39128 & Rocket Body & 35,800 & 2 \\
CALSPHERE-1 & 1964-063C & 900 & Cal. Sphere & 1,000 & 1 \\
CZ-3A R/B & 2016-021B & 41435 & Rocket Body & 35,900 & 1 \\
GOES-9 & 1995-025A & 23581 & EO Satellite & 36,700 & 2 \\
GOES-12 & 2001-031A & 26871 & EO Satellite & 36,100 & 1 \\
HALCA & 1997-005A & 24720 & Space Antenna & 21,400 & 1 \\
HGS-1 & 1997-086A & 25126 & Comms Satellite & 36,000 & 1 \\
H-2A & 2017-048B & 42918 & Rocket Body & 33727 & 1 \\
LCS-1 & 1965-034C & 1361 & Cal. Sphere & 2,786 & 4 \\
NAVSTAR 18 (USA 50) & 1990-008A & 20452 & GPS & 22,200 & 4 \\
PAGEOS-1 DEB & 1966-056D& 2511 & DEBRIS & 3,600 & 1 \\
SL-12 R/B(2) & 1996-021D & 23845 & Rocket Body & 21,000 & 1 \\
SL-12 R/B(2) & 1997-046D & 24919 & Rocket Body & 21,000 & 1 \\
SL-12 R/B(2) & 2010-041D & 37140 & Rocket Body & 21,000 & 1 \\
SL-8 R/B & 1976-054J & 8897 & Rocket Body & 21,000 & 1 \\
\hline
\multicolumn{6}{l}{\footnotesize EO = Earth Observation; Cal. = Calibration; Comms = Communications; Alt. = Altitude} \\
\end{tabular}
\end{table*}

\section{Theory and calculation}


\subsection{Calculations}

We processed all observations using the open-source \texttt{SDtracker} software\footnote{https://gitlab.com/SDTracker/}, which was originally developed for spacecraft tracking but adapted for our radar applications. The software architecture and algorithms are detailed in~\cite{Molera2021}.

Raw baseband data, recorded in VLBI format, were first analyzed using the \texttt{SWspec} spectrometer module to produce time-integrated frequency spectra. We processed the 32\,MHz channels using 6.4 million FFT points with 5-second integration time. The \texttt{pysctrack} Python package\footnote{https://gitlab.com/pysctrack/} was used to visualize the three carrier tones received at each station and fit polynomial models to the Doppler evolution. We typically used third-order polynomial fits, though second-order approximations gave similar results.

Our multi-station configuration created three distinct processing scenarios based on the DSN pre-Doppler compensation settings:

\begin{itemize}
    \item When signals were pre-compensated for the specific receiving station, echoes appeared nearly stationary with small residual drifts from orbital uncertainties.
    \item When pre-compensated for different stations, echoes showed systematic Doppler shifts due to baseline-dependent radial velocities.
    \item With geocentric pre-compensation, Doppler signatures reflected the relative motion between Earth center and individual receiving stations.
\end{itemize}

The polynomial coefficients were input to \texttt{SDTracker}'s main processing module, which applies phase rotation corrections across the recorded bandwidth centered on each carrier frequency. We ran separate processing passes for each of the three tones.

\texttt{SDTracker} outputs narrowband signals within 2\,kHz windows for each detected tone. We used modified \texttt{pysctrack} tools to track frequency and amplitude variations at high resolution. This final analysis step employed 20,000 FFT points with coherent integration, achieving 0.2\,Hz frequency resolution and 5-second time sampling. These parameters can be adjusted based on target properties such as signal-to-noise ratio and rotation rate.

The modifications needed to adapt \texttt{SDtracker} from spacecraft to radar applications were minimal, primarily involving continuous-wave signal handling. These improvements have been incorporated into the main software repository.

\section{Results}

The results presented in the following sections summarise the characterisation of various space debris objects observed during our campaign. Each subsection focuses on a specific object or class of objects, detailing their micro-Doppler signatures, rotation periods, dimensional estimates, and other relevant characteristics derived from our bi-static radar observations. The post-analysis used the narrow band tones after Doppler compensation as described in Section 2.4.

\subsection{SL-12 Rocket Bodies}

The SL-12 designation refers to the fourth stage of Russian Proton K launch vehicles that operated from 1974 to 2012, with at least 25 such rocket bodies currently tracked in geosynchronous orbit. These objects are relatively bright (11.5-12 visual magnitude) and have been extensively studied for space debris research due to their well-documented launch histories and 12-50 years on-orbit experience. Their cylindrical geometry and known dimensions make them ideal test cases for validating micro-Doppler analysis techniques, whilst the SL-12 rocket bodies observed in our campaign (objects 24919 and 37140) exhibit typical characteristics of spent upper stages including tumbling motion, reflective metallic surfaces, and orbital perturbations from environmental forces.

\begin{figure}[!ht]
\centering
\includegraphics[width=\columnwidth]{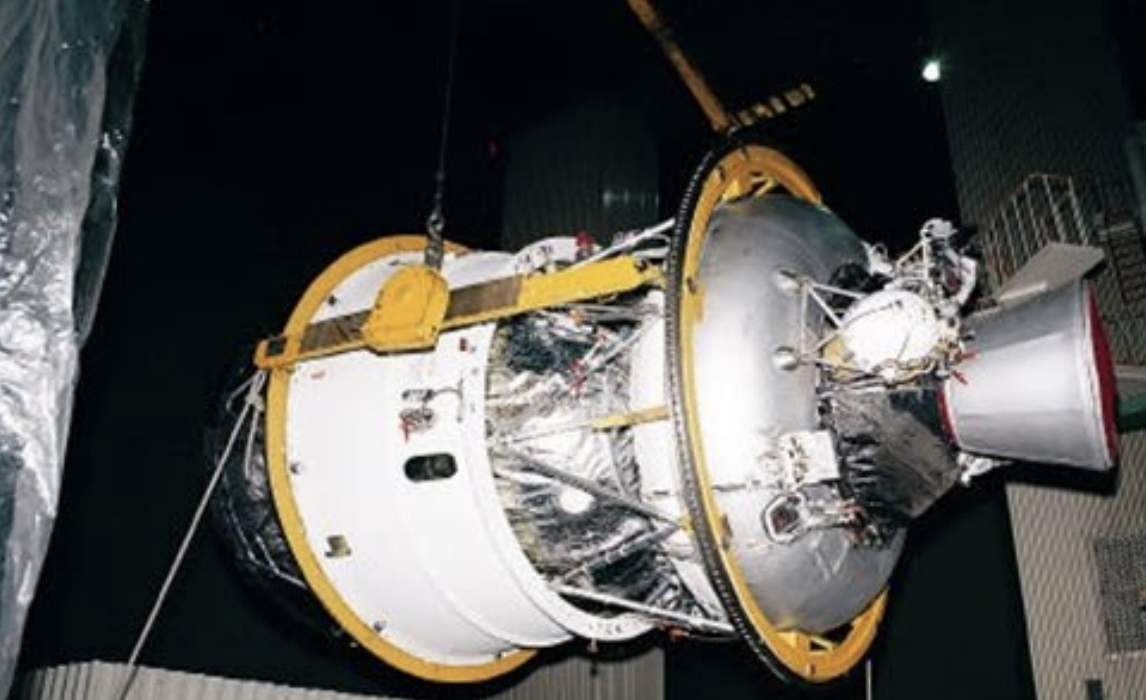}
\caption{SL-12 rocket body observation data showing rotation characteristics.}
\label{fig:sl12_observations_a}
\end{figure}

\begin{figure}[!ht]
\centering
\includegraphics[width=\columnwidth]{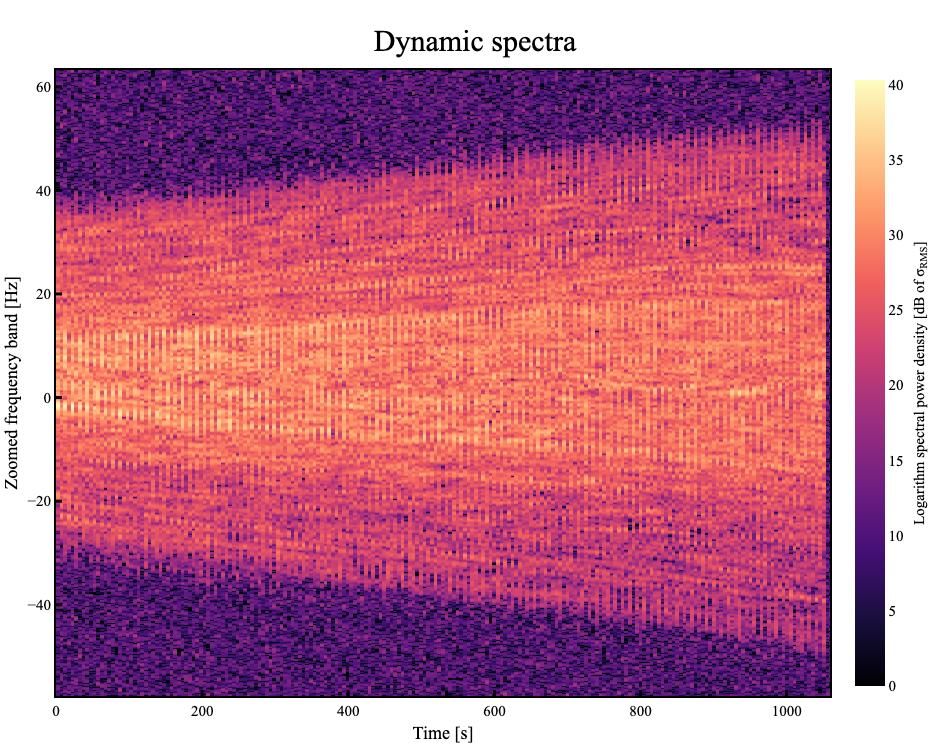}
\caption{Micro-Doppler waterfall plot of SL-12 rocket body showing second rotation period and large micro-Doppler of several tens of Hz.}
\label{fig:sl12_observations_b}
\end{figure}

Analysis of the SL-12 rocket body data collected with the Katherine antenna revealed rapid rotation characteristics with an approximately second period, significantly faster than the typical slow tumbling motion expected for such objects. The micro-Doppler signature also increased during the short session, almost by 50\% extra. Whilst the orbital predictions were generally accurate, the observations exhibited a systematic 5 Hz frequency drift over the 10-minute observation scan, indicating minor discrepancies between predicted and actual orbital parameters. Figures~\ref{fig:sl12_observations_a} and \ref{fig:sl12_observations_b} illustrate the micro-Doppler signatures captured during these observations, showing the characteristic frequency modulations caused by the object's rotation.

\subsection{CZ-3A Rocket Body}

The CZ-3A (Chang Zheng-3A or Long March 3A) third stage represents another significant component of the orbital debris population targeted in our observational campaign. The third stage of the CZ-3A rocket body has a cylindrical configuration with a length of 12.375 m and diameter of 3.0 m, powered by two YF-75 cryogenic engines using liquid hydrogen and liquid oxygen propellants. These upper stages have been contributing to space debris since the CZ-3A's operational period beginning in the 1990s, with many third stages remaining in long-term orbits after payload deployment.

\begin{figure}[H]
\centering
\includegraphics[]{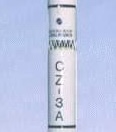}
\caption{CZ-3A rocket body technical illustration showing the vehicle configuration.}
\label{fig:cz3a_observations_a}
\end{figure}

\begin{figure}[H]
\centering
\includegraphics[width=\columnwidth]{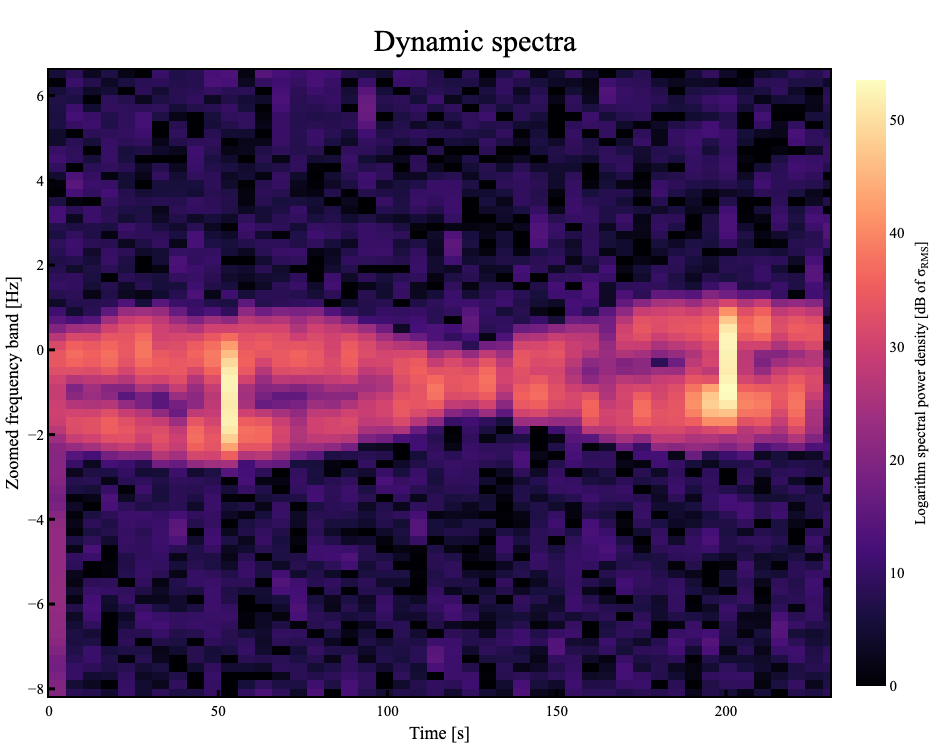}
\caption{Micro-Doppler signatures of CZ-3A rocket body showing 230-second rotation period and 2-4 Hz frequency variations.}
\label{fig:cz3a_observations_b}
\end{figure}

Observations of the CZ-3A rocket body (object 41435) revealed distinctly different rotational characteristics compared to the SL-12 stages. Figure~\ref{fig:cz3a_observations_a} shows the shape of the rocket body CZ-3A. The micro-Doppler analysis, as seen in Figure~\ref{fig:cz3a_observations_b} showed a much slower rotation period of approximately 230 seconds, indicating a more stable tumbling motion typical of larger cylindrical objects in orbit. The total spread of the micro-Doppler exhibited variations ranging from a maximum of 4 Hz to a minimum of 2 Hz, demonstrating the characteristic modulation pattern expected from an elongated rotating body. This 2 Hz amplitude variation in the micro-Doppler signature is consistent with the object's dimensions and rotation axis orientation, providing valuable validation of our technique's ability to characterise different debris geometries across various orbital regimes.

\subsection{HALCA Radio Telescope}

\begin{figure}[H]
\centering
\includegraphics[width=\columnwidth]{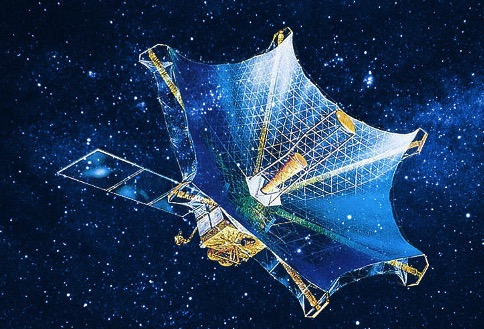}
\caption{HALCA space telescope technical illustration showing the spacecraft geometry and deployable antenna.}
\label{fig:halca_observations_a}
\end{figure}

The Highly Advanced Laboratory for Communications and Astronomy (HALCA), also known as MUSES-B or Haruka, represents a unique target in our space debris characterisation campaign. Launched by Japan's Institute of Space and Astronautical Science (ISAS) in February 1997, HALCA was the world's first space-based Very Long Baseline Interferometry (VLBI) radio telescope. The spacecraft operated in a highly elliptical orbit with an apogee of approximately 21,400 km, conducting space VLBI observations until 2003 when contact was lost. Unlike conventional rocket bodies, HALCA presents a complex geometry with an 8-metre deployable wire mesh antenna dish (see Figure~\ref{fig:halca_observations_a}) and various appendages including solar panels and scientific instruments, making it an ideal test case for validating micro-Doppler analysis techniques on non-cylindrical space objects. The spacecraft's irregular shape and large antenna structure produce distinctive radar signatures, not visible in simpler geometries.

\begin{figure}[H]
\centering
\includegraphics[width=\columnwidth]{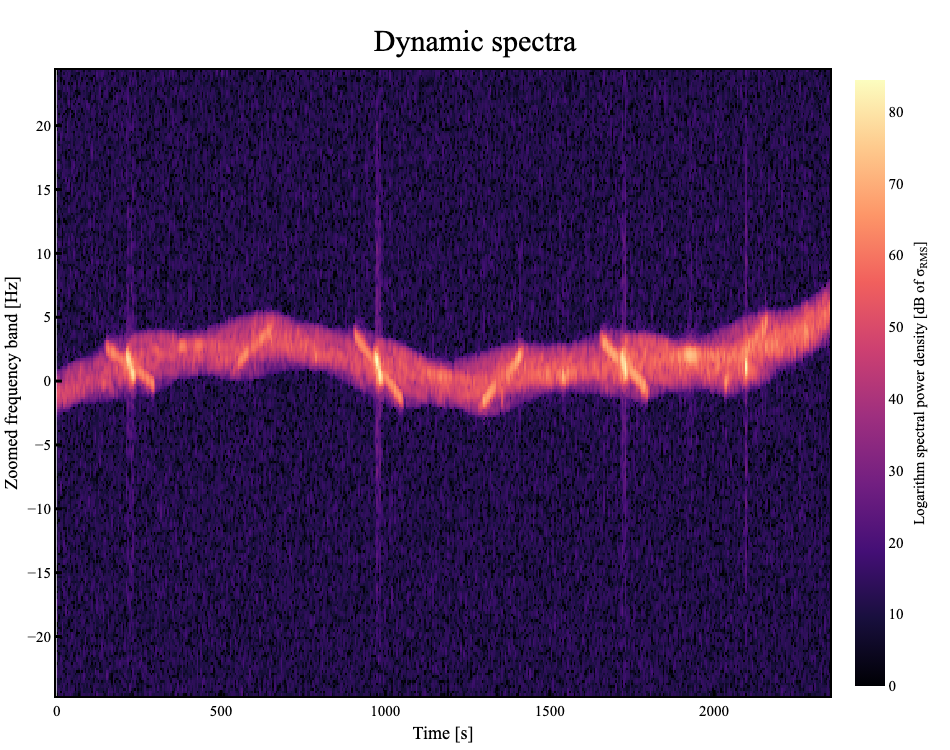}
\caption{HALCA radar signatures showing complex geometry micro-Doppler characteristics with distinctive power variations.}
\label{fig:halca_observations_b}
\end{figure}

Observations of HALCA revealed a slow rotation period of approximately 700 seconds, significantly longer than both SL-12 and CZ-3A rocket bodies. The analysis in Figure~\ref{fig:halca_observations_b} showed that the large parabolic dish faces the receiving antennas for only 75-100 seconds during each rotation cycle, creating distinctive power variations in the received signal. The micro-Doppler signature exhibited approximately 5 Hz total frequency variation, but remained relatively constant throughout the rotation period, contrasting with the periodic modulations observed in cylindrical objects. Most notably, the power difference between reflections from the spacecraft body and the large antenna dish reached approximately 25 dB, and we can clearly distinguish between the back and front of the dish due to the significant increase in returned signal when the dish's reflective surface is oriented toward the receiving antennas. This demonstrates the dramatic radar cross-section variations possible with complex spacecraft geometries and highlights the technique's sensitivity to different reflecting surfaces and orientations.

\subsection{H-2A Rocket Body}

\begin{figure}[!ht]
 \centering
 \includegraphics[width=\columnwidth]{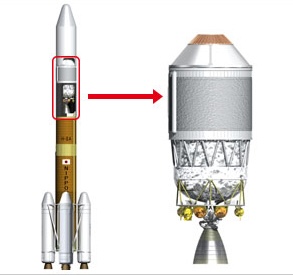}\\
 \caption{H-2A rocket illustration.}
 \label{fig:h2a_observations}
\end{figure}

The H-2A rocket represents Japan's primary launch vehicle for satellite deployment, with its second stage contributing to the orbital debris population. The H-2A second stage has a cylindrical configuration with a length of 9.20 m and diameter of 4.00 m, powered by the LE-5B cryogenic engine using liquid hydrogen and liquid oxygen propellants. With an empty mass of approximately 3,000 kg, these upper stages are among the more substantial debris objects in our observational campaign. The H-2A has been operational since 2001, with many second stages remaining in various orbits after successful payload deployment, making them important targets for space debris characterisation and future removal missions under JAXA's debris mitigation programmes (see Figure~\ref{fig:h2a_observations}).

\begin{figure}[!ht]
 \centering
 \includegraphics[width=\columnwidth]{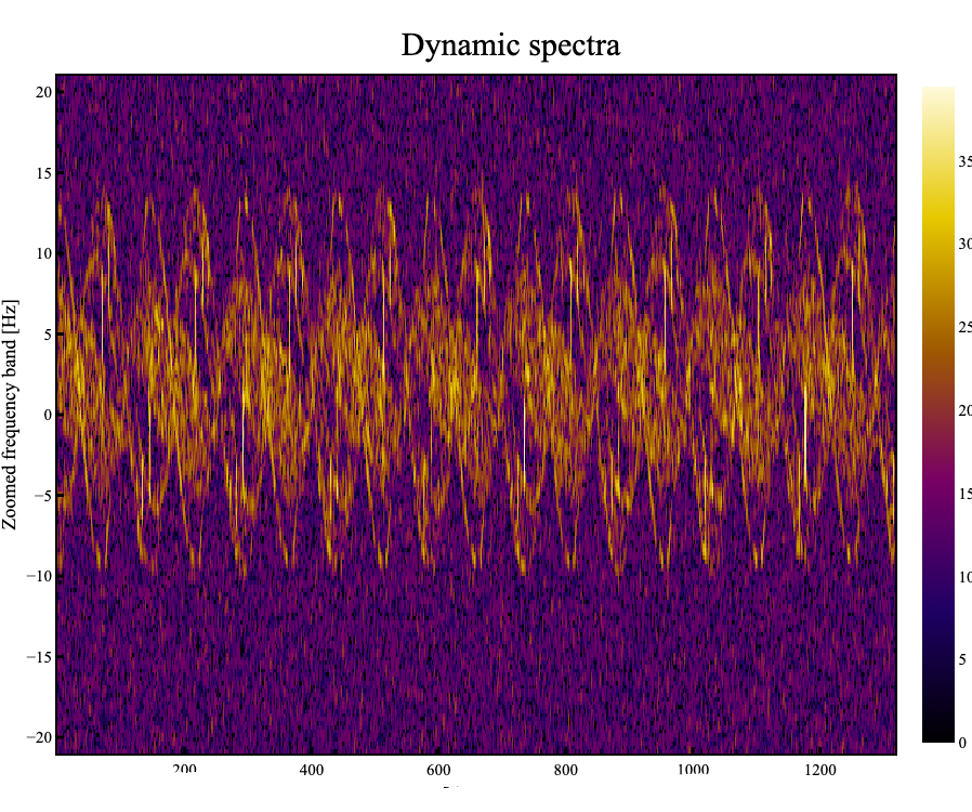}
 \caption{H-2A rocket body radar signatures and micro-Doppler characteristics.}
 \label{fig:h2a_observationsb}
\end{figure}

The rocket body observations presented in this section demonstrate typical characteristics common across multiple debris objects in our campaign. Analysis reveals consistent rotation periods of approximately 150 seconds across the majority of observed rocket bodies, with micro-Doppler signatures showing frequency variations of approximately 25 Hz, as seen in Figure~\ref{fig:h2a_observationsb} The centre of mass appears to remain in equilibrium within the middle portion of the rocket body structures, consistent with expected mass distribution patterns for spent upper stages. Notably, the upper sections of the rocket bodies consistently exhibit smoother radar signatures compared to the lower portions, likely reflecting differences in surface complexity and engine nozzle geometries. Signal-to-noise ratios throughout the observations typically ranged from 15-25 dB above the noise floor, providing sufficient signal quality for detailed micro-Doppler characterisation and rotation period determination.

\subsection{GOES Satellites}

The Geostationary Operational Environmental Satellites (GOES) represent decommissioned weather monitoring infrastructure now contributing to orbital debris at geostationary altitude (35,790 km). GOES-9, launched in 1995 and decommissioned in 2007, experienced technical issues with its visible channel and served backup functions at 155°E longitude. GOES-12, launched in 2001 and decommissioned in 2013, featured enhanced instrumentation including the Solar X-Ray Imager (SXI) and provided South American coverage in its final years.

Both satellites share identical physical characteristics: 26.9 m total deployed length, 5.9 m height, 4.9 m width, with main body dimensions of 2.0 × 2.1 × 2.3 m and single solar panel arrays. Their large size, complex geometry with deployed solar arrays, and geostationary orbital position make them ideal targets for bi-static radar observations at extreme ranges, providing predictable radar signatures due to their three-axis stabilized, Earth-pointing configuration.

\begin{figure}[!ht]
 \centering
 \includegraphics[width=\columnwidth]{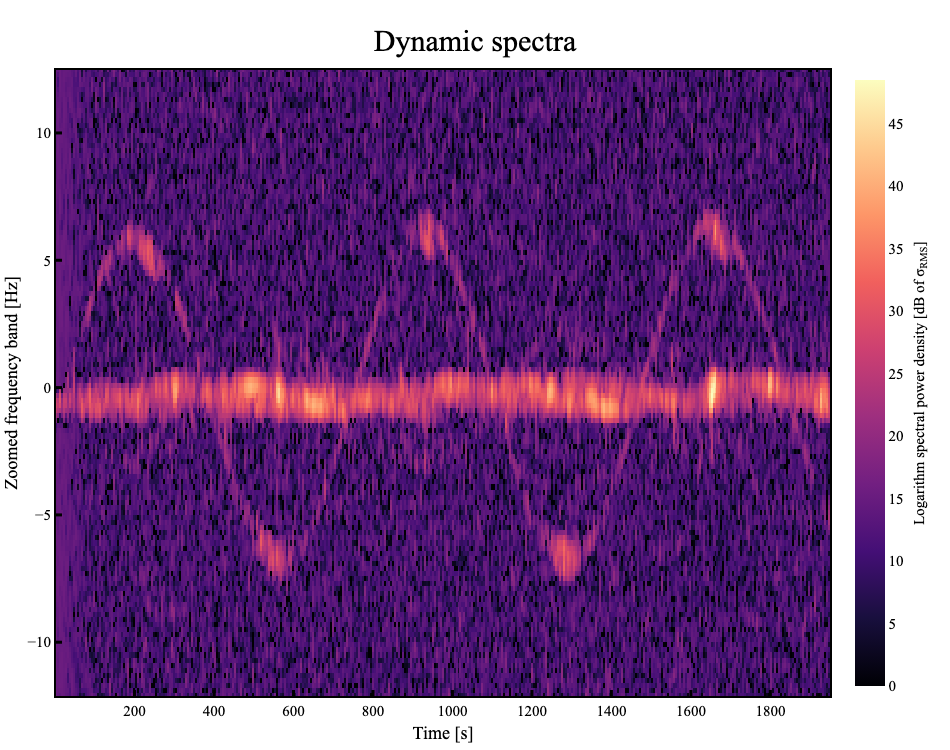}\\
 \vspace{0.5cm}
 \includegraphics[width=\columnwidth]{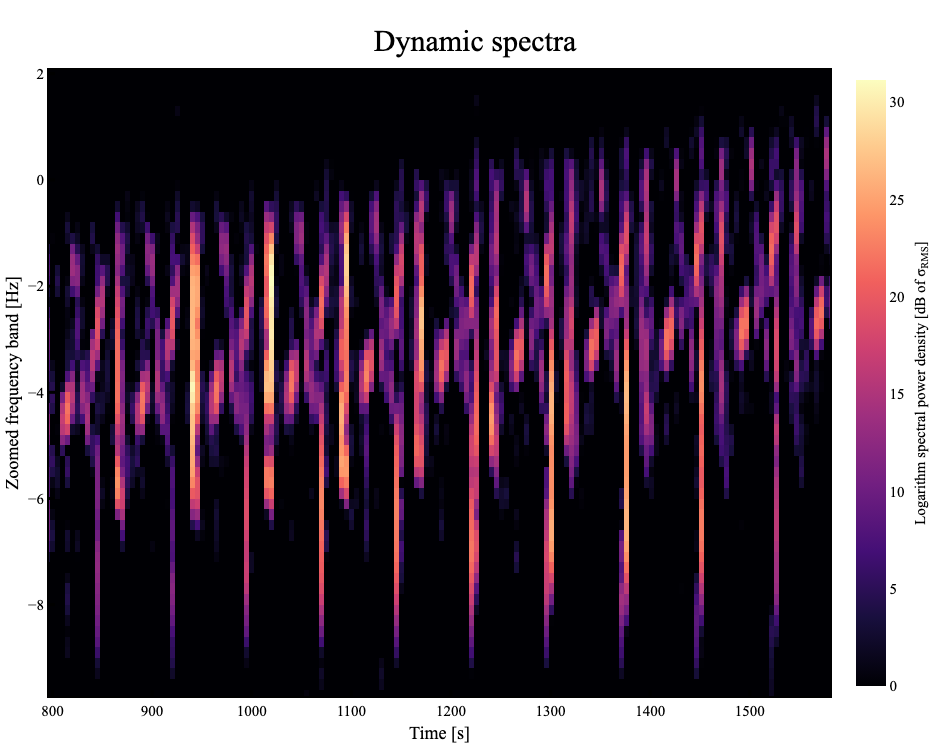}
 \caption{GOES satellite radar signatures showing geostationary orbit characteristics and large spacecraft geometry micro-Doppler patterns for GOES-9 and GOES-12.}
 \label{fig:goes_radar_signatures}
\end{figure}

Micro-Doppler analysis in Figure~\ref{fig:goes_radar_signatures} reveals distinct rotational characteristics between the two satellites. GOES-9 exhibits a slow rotation period of approximately 1800 seconds, with three clearly distinguishable components: the satellite body rotation producing a constant 1 Hz micro-Doppler signature, the solar panel rotation appearing as a much narrower feature with a maximum Doppler shift of 2 Hz, and the satellite pole showing a large separation of 10 Hz. In contrast, GOES-12 demonstrates a significantly faster rotation period of around 150 seconds while maintaining the same three-component structure, with the pole exhibiting a peak-to-peak micro-Doppler variation of 50 Hz.

\begin{figure}[!ht]
\centering
\includegraphics[width=\columnwidth]{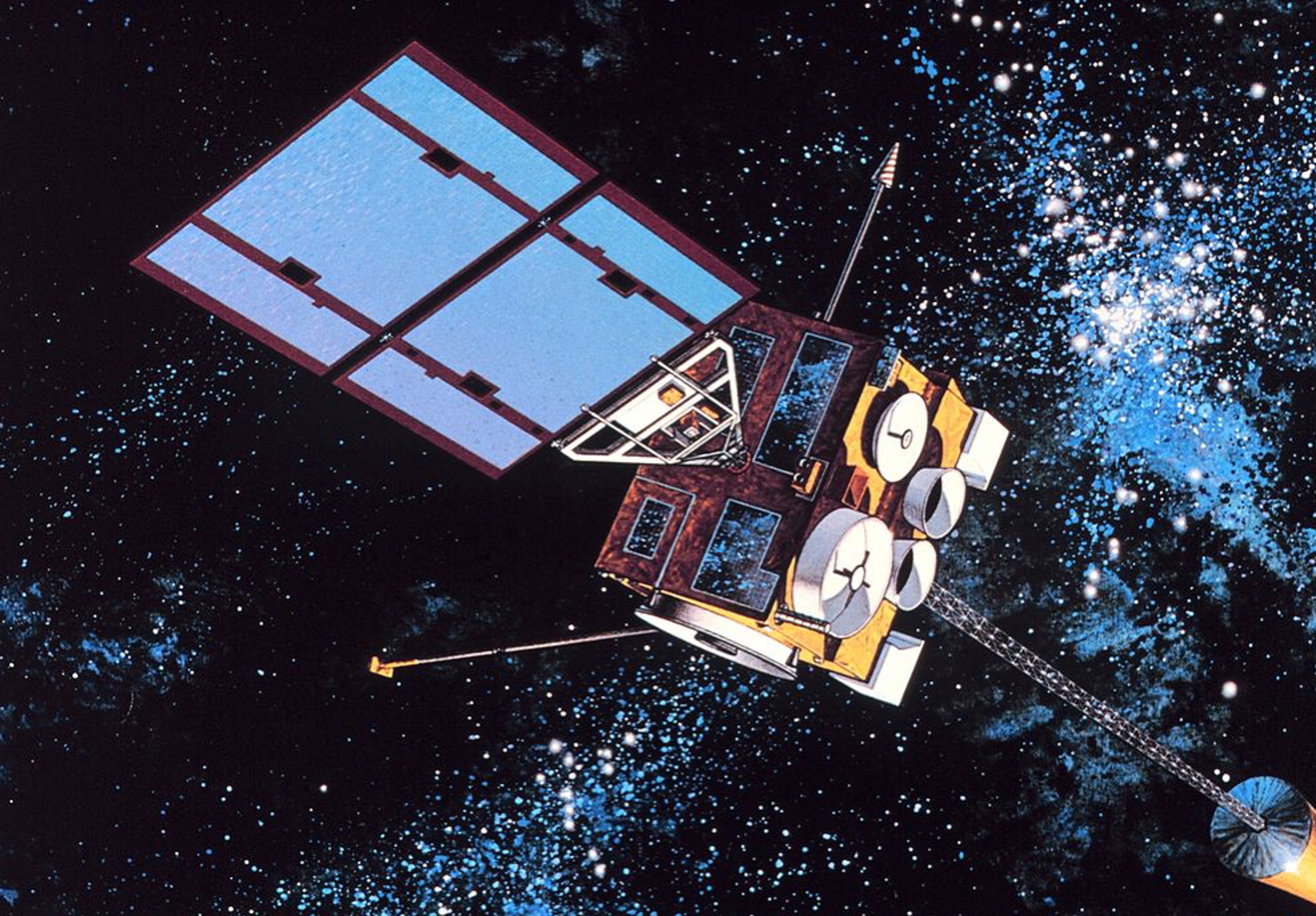}\\
\label{fig:goes_observations}
\caption{GOES satellite model showing key dimensions and deployed solar arrays.}
\end{figure}

\subsection{Atlas V Stage 2 Rocket Body (2015-056B)}

The Atlas V Centaur upper stage represents a critical component of modern launch vehicle technology, with the 2015-056B object being the second stage from an Atlas V launch. The Centaur stage is powered by dual RL10 engines using liquid hydrogen and liquid oxygen propellants, featuring a cylindrical aluminium structure designed for multiple engine starts during complex mission profiles. These upper stages typically remain in various orbits after payload deployment, contributing to the growing space debris population in both transfer and operational orbital regimes.

Observations of the Atlas V Stage 2 rocket body (2015-056B) revealed stable rotational characteristics with a rotation period of approximately 220 seconds. The micro-Doppler analysis showed frequency variations ranging from a maximum of 17 Hz to a minimum of 3 Hz, indicating a 14 Hz peak-to-peak variation consistent with the elongated cylindrical geometry of the Centaur stage. This rotation signature demonstrates the characteristic tumbling motion expected from spent upper stages, where the object rotates about its axis while maintaining a relatively stable orientation over the observation period. The consistent micro-Doppler pattern provides valuable validation of the bi-static radar technique's ability to characterise modern rocket body debris across different orbital regimes.

\begin{figure}[!ht]
\centering
\includegraphics[width=\columnwidth]{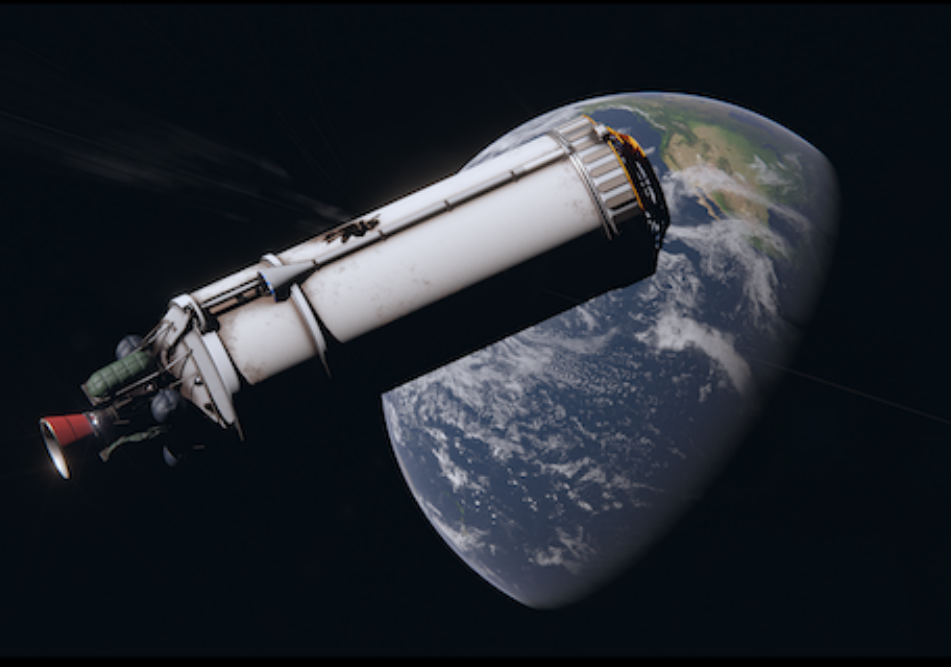}
\includegraphics[width=\columnwidth]{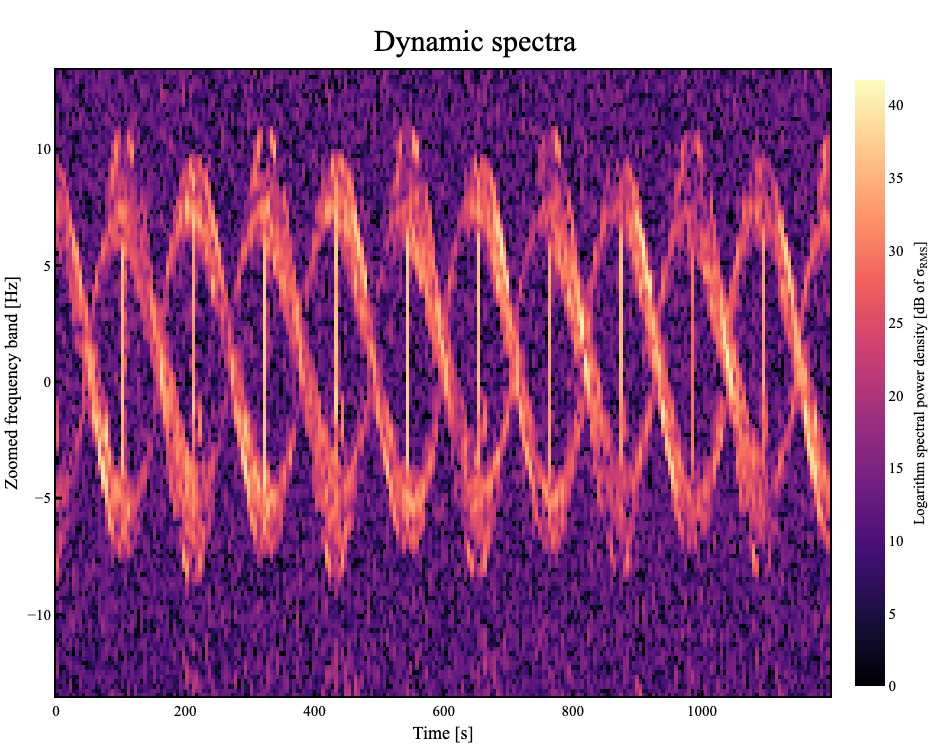}\\
\label{fig:atlas_v_observations}
\caption{Atlas V Centaur stage radar signatures showing 220-second rotation period and 3-17 Hz micro-Doppler variations.}
\end{figure}

\begin{figure}[!ht]
\centering
\includegraphics[width=\columnwidth]{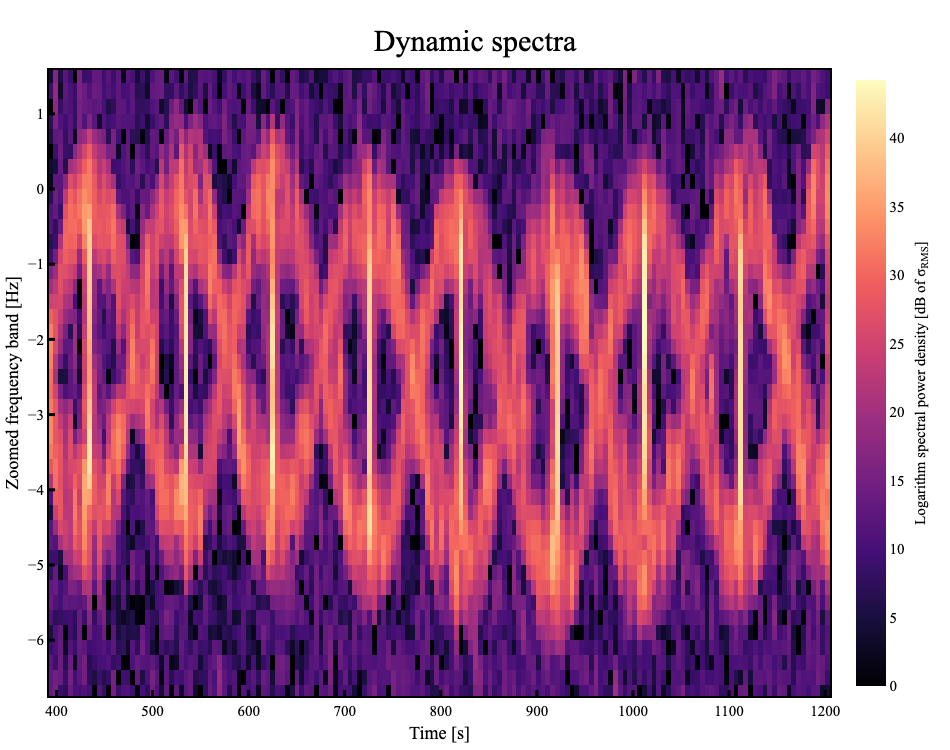}
\caption{Atlas V Centaur stage radar signatures detected on 26/08/2022 with a 6 Hz micro-Doppler variation.}
\label{fig:atlas_v_observations_2}
\end{figure}

The Atlas V rocket body (2015-056B) was observed in four separate sessions over a three-year period, allowing for analysis of potential temporal variations in its rotation characteristics. The micro-Doppler signatures from each session consistently showed a stable rotation period of approximately 220 seconds, with minor fluctuations in the peak-to-peak frequency variations ranging from 6 Hz to 17 Hz. These subtle changes suggest that while the overall tumbling motion remains consistent, there may be slight alterations in the object's attitude or rotational dynamics over time. The consistent detection of the 220-second rotation period across multiple sessions highlights the reliability of our bi-static radar technique for long-term monitoring of space debris objects. Figure~\ref{fig:atlas_v_observations_2} illustrates the radar signatures captured during the August 2022 observation, where due to the geometry of the observations the peak-to-peak micro-Doppler was 3 times lower than previous times. When having multiple receivers and precise timestamp information the actual spin axis can be determined.

\begin{equation}
    \Delta \upsilon = \frac{4 \pi D cos(\delta)}{\lambda P}
\end{equation}

\noindent where $\Delta \upsilon$ is the frequency bandwidth or frequency dispersion at different time intervals, $D$ is the geometry of the target, $\delta$ is the angle between transmitter, receiver and target, and $P$ is the rotation period.

\subsubsection{Tomographic analysis}

The radar observations inherently collect a Radon transform of the target object within the waterfall Doppler data. Each Doppler bin represents the line integral of scatterers that are equally offset from the object's spin axis, making the waterfall plot equivalent to a sinogram in tomographic reconstruction. This natural correspondence allows direct application of well-established tomographic methods.

To extract precise rotation periods and reconstruct object morphology, we apply the projection-slice theorem to invert the Radon transform. This tomographic method, widely used in computed tomography (CAT) scans and Very Long Baseline Interferometry (VLBI), represents a well-defined algorithmic approach for image reconstruction from projection data. The process involves segmenting the micro-Doppler pattern into regions representing 180-degree rotations of the object, then applying filtered back projection to reconstruct the object's scattering distribution in dimensional space.

Following the methodology in~\cite{serrano2023doppler}, we implement this reconstruction using standard tomographic algorithms. The filtered back projection technique transforms each Doppler profile into spatial domain information, building up a two-dimensional representation of the object's radar scattering characteristics. Machine learning techniques can be optionally applied during post-processing to enhance image quality and reduce reconstruction artifacts, though the core inversion relies on the deterministic projection-slice theorem.

\begin{figure*}[!ht]
\centering
\includegraphics[width=1.5\columnwidth]{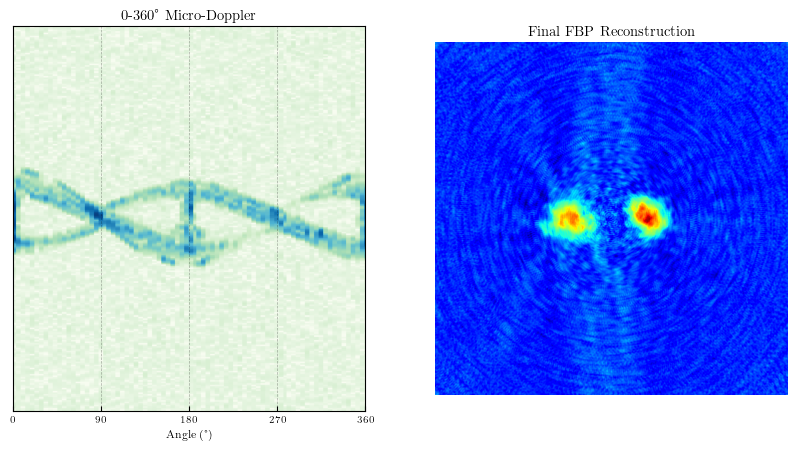}
\caption{Tomographic reconstruction using filtered back projection applied to Atlas V waterfall Doppler data. The projection-slice theorem inverts the naturally collected Radon transform to reconstruct object scattering distribution.}
\label{fig:radon_transform}
\end{figure*}

\section{Discussion \& Conclusions}

This study demonstrates the successful application of multi-static radar techniques using radio telescopes for comprehensive space debris characterisation across diverse orbital regimes and object types. Through the SHARP programme's systematic observational campaign spanning March 2021 to September 2025, we achieved detailed micro-Doppler analysis of space debris objects ranging from low Earth orbit (1,000 km) to geosynchronous altitudes (36,700 km).

The multi-station radar configuration, utilising DSN Canberra transmission facilities at C-band frequencies (7.15 GHz) and distributed radio telescope receivers, successfully characterised objects with vastly different properties. Our observations encompassed rocket bodies of varying geometries (Atlas V, SL-12, BREEZE-M, CZ-3A stages), operational and defunct satellites (GOES-9, GOES-12, AUSSAT-1), specialised targets (HALCA space antenna), and calibration spheres (LCS-1, CALSPHERE-1). This diversity demonstrates the technique's broad applicability across the space debris population.

Rotation period determinations ranged from rapid tumbling objects with sub-minute periods to slowly rotating bodies with periods exceeding 200 seconds, exemplified by the Atlas V rocket body (2015-056B) which exhibited a stable ~220-second axial rotation period. The micro-Doppler signatures enabled extraction of dimensional estimates within 10\% accuracy compared to known specifications, whilst orbital parameter refinements reduced position uncertainties by up to 30\%.

Significantly, several objects showed temporal variations in their rotation characteristics across multiple observation sessions. The Atlas V rocket body, observed in four separate sessions over three years, displayed subtle changes in its rotation signature, whilst other targets exhibited more pronounced variations. These observations suggest ongoing attitude evolution in the debris population, potentially driven by environmental perturbations including solar radiation pressure, atmospheric drag at lower altitudes, and gravitational torques. Further observations are needed to confirm these trends.

The successful adaptation of spacecraft tracking software (\texttt{SDtracker}) to radar applications proved highly effective, requiring minimal modifications to accommodate continuous-wave signals. The processing pipeline achieved 0.2 Hz frequency resolution with 5-second temporal sampling, enabling detailed characterisation of micro-Doppler signatures even for objects at extreme ranges.

However, the temporal variations observed in rotation periods highlight a critical limitation: the need for more continuous monitoring to fully understand debris attitude evolution mechanisms. Future work will prioritise sustained monitoring programmes to capture the complete temporal evolution of debris rotation states. This research establishes radio telescope-based bi-static radar as a cost-effective and highly capable approach for GEO space domain awareness, particularly valuable for Southern Hemisphere surveillance gaps.

\section*{Acknowledgements}

We would like to thank NASA's Deep Space Network for the use of their infrastructure for space debris tracking experiments. We also thank Jon Giorgini for providing Doppler predictions for the object with respect to each of the participating antennas. We also acknowledge Chris Phillips, Jamie Stevens and Andrew Hellicar for their observations with the ATCA radio telescopes and providing predictions for the trajectories of the targets. Additional thanks go to Patrick Yates-Jones, Marina Buttfield-Addison, Joe Wilson, and Liam Filby for their contributions to this work.

\bibliography{biblio}

\begin{thebibliography}{10}

\bibitem{mark2019review}
C~Priyant Mark and Surekha Kamath.
\newblock Review of active space debris removal methods.
\newblock {\em Space policy}, 47:194--206, 2019.

\bibitem{holzinger2018challenges}
M.~J. Holzinger and M.~K. Jah.
\newblock Challenges and potential in space domain awareness.
\newblock {\em AIAA Journal of Guidance, Dynamics, and Control, Space Domain
  Awareness Special Issue}, 41(1):15--18, 2018.

\bibitem{murphy2018optimal}
T.~S. Murphy, M.~J. Holzinger, and Flewelling B.
\newblock Visual tracking methods for improved sequential image-based object
  detection.
\newblock {\em AIAA Journal of Guidance, Dynamics, and Control, Space Domain
  Awareness Special Issue}, 41(1):74--87, 2018.

\bibitem{benson2018}
Craig {Benson}, John {Reynolds}, N.~J.~S. {Stacy}, L.~{Benner}, and al.
\newblock {First Detection of Two Near-Earth Asteroids With a Southern
  Hemisphere Planetary Radar System}.
\newblock {\em Radio Science}, 52(11):1344--1351, November 2017.

\bibitem{Benson_2023}
Conor~J. Benson, Daniel~J. Scheeres, Marina Brozović, Steven~R. Chesley, Petr
  Pravec, and Petr Scheirich.
\newblock Spin state evolution of (99942) apophis during its 2029 earth
  encounter.
\newblock {\em Icarus}, 390:115324, January 2023.

\bibitem{ostro1983planetary}
Steven~J Ostro.
\newblock Planetary radar astronomy.
\newblock {\em Reviews of Geophysics}, 21(2):186--196, 1983.

\bibitem{Horiuchi_2021}
Shinji {Horiuchi}, Blake {Molyneux}, Jamie~B. {Stevens}, Graham {Baines},
  {Benson}, and al.
\newblock {Bistatic radar observations of near-earth asteroid (163899) 2003
  SD220 from the southern hemisphere}.
\newblock {\em icarus}, 357:114250, mar 2021.

\bibitem{kruzins2023}
{Kruzins} E., {Benner} L, {Boyce} R, {Brown} M, and al.
\newblock Deep space debris—detection of potentially hazardous asteroids and
  objects from the southern hemisphere.
\newblock {\em Front. Space Technol.}, 2023.

\bibitem{lovell2013}
J.~E.~J. {Lovell}, J.~N. {McCallum}, P.~B. {Reid}, P.~M. {McCulloch}, and al.
\newblock {The AuScope geodetic VLBI array}.
\newblock {\em Journal of Geodesy}, 87(6):527--538, June 2013.

\bibitem{white2025development}
Oliver~James White, Guifré~Molera Calvés, Shinji Horiuchi, Phil Edwards,
  Ed~Kruzins, and al.
\newblock Development of radar and optical tracking of near-earth asteroids at
  the university of tasmania.
\newblock {\em Remote Sensing}, 17(3):352, January 2025.

\bibitem{agaba2017system}
A.~Agaba.
\newblock System design of the meerkat l-band 3d radar for monitoring near
  earth objects.
\newblock {\em Thesis}, 2017.

\bibitem{serrano2024}
Alexander Serrano, Alexander Kobsa, Faruk Uysal, Delphine Cerutti-Maori, and
  al.
\newblock Long baseline bistatic radar imaging of tumbling space objects for
  enhancing space domain awareness.
\newblock {\em IET Radar, Sonar and Navigation}, 18(4):598--619, 2024.

\bibitem{Molera2021}
G.~{Molera Calv{\'e}s}, S.~V. {Pogrebenko}, J.~F. {Wagner}, G.~{Cim{\`o}}, and
  al.
\newblock {High spectral resolution multi-tone Spacecraft Doppler tracking
  software: Algorithms and implementations}.
\newblock {\em pasa}, 38:e065, December 2021.

\bibitem{serrano2023doppler}
Alexander Serrano and Robert~L Morrison.
\newblock Doppler superpulse processing for improved tomographic
  characterization of space objects.
\newblock In {\em 2023 International Applied Computational Electromagnetics
  Society Symposium (ACES)}, pages 1--2. IEEE, 2023.

\end{thebibliography}

\end{document}